\documentclass[lettersize,journal]{IEEEtran}
\usepackage{cite}
\usepackage{amsmath,amssymb,amsfonts}
\usepackage{graphicx}
\usepackage{textcomp}
\usepackage{xcolor}

\usepackage{tikz}
\usepackage{amsmath}
\usepackage{float}
\usepackage{subfig}

\usepackage{amssymb,pifont}
\usepackage{array,colortbl,multirow,multicol,booktabs}
\usepackage{threeparttable}

\usepackage{algorithm}  
\usepackage{algpseudocode}  
\begin{document}

\title{Chorusing Synchronization Signals for Ambient \\ 5G Backscatter}

\author{Yunyun Feng,~\IEEEmembership{Student Member,~IEEE,}
        Chenhong Cao,~\IEEEmembership{Member,~IEEE,}
        Si Chen,~\IEEEmembership{Student Member,~IEEE}
        and Wei Gong,~\IEEEmembership{Senior Member,~IEEE}
\IEEEcompsocitemizethanks{\IEEEcompsocthanksitem Yunyun Feng, Chenhong Cao, and Wei Gong are with the School of Computer Science and Technology, University of Science and Technology of China, Hefei 230026, China (E-mail: yunyunf@mail.ustc.edu.cn, chcao@ustc.edu.cn, weigong@ustc.edu.cn).\protect

        Si Chen is with the School of Computing Science, Simon Fraser University, Burnaby BC V5A 1S6, Canada (E-mail: sca228@sfu.ca).\protect
        
        (Corresponding author: Wei Gong.)
	}
 }




\maketitle

\begin{abstract}
5G backscatter communication presents an emerging energy-efficient IoT connectivity solution with enhanced availability and data rate advantages over traditional wireless networks. 
    For 5G backscatter, synchronization is crucial as it ensures high-quality transmission.
    Popular synchronization methods employ autocorrelation and cross-correlation for accurate timing, yet they are constrained by resources. 
    Traditional cross-correlation-based methods for resource utilization optimization also fail in 5G backscatter due to the presence of multiple templates for 5G.
    A synchronization strategy that supports high accuracy and low power would be highly attractive for wireless backscatter communication. 
    We propose Symmetric Differential (SD)-based Sync, an accurate and resource-efficient synchronization method for 5G backscatter.  
    We have observed that the envelope of the 5G Primary Synchronization Signal (PSS) exhibits a unique mirror symmetry, which enables us to employ differential techniques for low-power PSS detection.
    We extensively evaluated our design using a testbed of backscatter hardware, SDR gNodeB, and User Equipment (UE). 
    Results show that our SD consumes 3,175 D flip-flops, which is 87x lower than NR fine timing (NFT), 181x lower than symmetry-based semi-template sync (SST), and 30x lower than symmetric autocorrelation (SA)-based sync.
\end{abstract}

\begin{IEEEkeywords}
Backscatter communication, Cellular network, Internet-of-Things.
\end{IEEEkeywords}

\section{Introduction}
    \IEEEPARstart{T}{he} rapid growth of the IoT demands scalable, energy-efficient connectivity, for which backscatter communication offers a compelling ultra-low-power solution \cite{lu2023millimeter}.
    A typical ambient backscatter system involves three key steps: an ambient transmitter provides the carrier signal; the tag synchronizes with it to align frames and transmit sensor data; and a receiver captures the backscattered signal to recover the data.
    However, traditional ambient backscatter struggles to balance accessibility with high data rates.
    5G New Radio (NR) backscatter emerges as a promising alternative, offering higher data rates through wide bandwidth, greater connectivity via frequency and spatial diversity, and improved accessibility through continuous coverage.  
    An illustrative application of 5G backscatter is low-power video streaming for remote surveillance. Backscatter-enabled cameras transmit video by reflecting and modulating ambient 5G signals, eliminating the need for battery-powered transmitters. Leveraging 5G’s high throughput and low latency, the system enables real-time, high-quality streaming in power-limited or hard-to-access environments, significantly reducing energy consumption and maintenance costs.
    
Synchronization is vital in 5G backscatter systems; without it, tags would modulate data randomly onto incoming carriers, resulting in increased inter-symbol interference (ISI) and reduced throughput \cite{dunna2021syncscatter}.
Extensive research has focused on high-accuracy synchronization algorithms, typically combining autocorrelation and cross-correlation for timing and frequency alignment \cite{nasraoui2014simply,wang2021novel}.
This stems from their complementary characteristics: autocorrelation is resilient to frequency offset but vulnerable to noise, while cross-correlation is more robust to noise but sensitive to frequency shifts \cite{zhang2017detecting,huang2012joint}.
These algorithms, however, are energy-intensive and suited for resource-rich active radios, making them impractical for the constrained environments of passive radios \cite{peng2018plora}.
Moreover, passive radios designed for low-power IoT applications typically operate at much lower data rates than active radios \cite{stanacevic2020backscatter}, and thus require only microsecond-level synchronization accuracy instead of nanosecond-level precision.

To save power, backscatter radios rely on passive envelope detectors to extract the carrier envelope, rather than processing IQ samples, which diminishes the frequency offset tolerance of autocorrelation.
As a result, systems like PLoRa \cite{peng2018plora} and SyncLTE \cite{feng2023heartbeating} adopt cross-correlation for synchronization.
However, conventional cross-correlation with one template cannot handle 5G signals with varying cell IDs, as there are three distinct Primary Synchronization Signal (PSS) sequences.
To address this, synchronization must be performed using three predefined templates.
This naturally leads us to consider 5G NR Fine Timing (NFT), which cross-correlates the received signal with all three PSS templates to locate the correlation peak and identify the PSS.
However, this method imposes significant computational overhead, far beyond the capabilities of low-power backscatter tags.

    Given cross-correlation's popularity in active sync, we explore its feasibility in passive radio. Thus, our question is:\\
    \textbf{Q1: Is cross-correlation-based synchronization applicable to 5G backscatter?}

    Cross-correlation, renowned for its robustness to noise, is a widely adopted synchronization technique in active radios, enabling fine symbol alignment in systems such as WiFi, LTE, and 5G.
    While it can provide high synchronization accuracy in backscatter systems, its computational intensity poses a major challenge for 5G backscatter applications.
    As previously noted, 5G NR Fine Timing (NFT) consumes substantial resources, motivating our exploration of ways to optimize cross-correlation for low-power tags.
    The total computational burden of cross-correlation grows with both the template size and the number of correlation operations.
    In 5G, this burden is amplified by the large size of the PSS-based templates, which span tens of microseconds.
    To address this, we observe that the 5G PSS envelope exhibits mirror symmetry.
    Leveraging this property, we propose Symmetry-based Semi-Template synchronization (SST).
    SST first employs autocorrelation to exploit the symmetry and reduce correlation operations, followed by cross-correlation using only half of the PSS template for peak detection.
    Although SST significantly reduces resource consumption compared to NFT, it still exceeds the capabilities of typical backscatter tags.
    Even after lowering the sampling rate to shrink template size, we find no viable trade-off between accuracy and computational cost.

    Since continuing along the path of cross-correlation-based synchronization is unfeasible, the second question emerges: 
    \textbf{Q2: How can we design ultra-low-power and accurate synchronization for 5G backscatter?}

    We further observe that the 5G PSS envelope uniquely exhibits mirror symmetry, which motivates its exploitation for PSS detection.
    A direct synchronization approach, symmetric autocorrelation (SA)-based sync from \cite{zhang2012low}, uses autocorrelation to confirm symmetry and detect the central position of PSS, which is also the first step of SST. 
    However, due to the higher resource consumption of multipliers than adders in autocorrelation, SA fails to meet the requirements of low-power tags. Therefore, we exploit the property of envelope equality at symmetrical points to \textit{transform multiplications into additions}, significantly reducing computational resources. Moreover, we propose symmetric differential (SD)-based sync, which employs differential techniques to locate the minimum value for PSS detection. It achieves a substantial reduction in computational resources without compromising synchronization performance, thus meeting the low-power synchronization requirements for 5G backscatter.

    To validate the feasibility of our design, we implemented SD in MATLAB and on a low-power FPGA. Through extensive experiments, we show that
    \begin{itemize}
    	\item SD consumes the fewest resources among NFT, SST, and SA, well below the capacity of the AGLN250 FPGA, whereas other methods far exceed this capacity. It consumes 87× fewer resources than NFT, 181× fewer than SST, and 30× fewer than SA. 
            \item IC simulation power of SD with quantization is 70.06 \textmu W, which is 4.3x lower than SD without quantization. Moreover, the median sync errors for SD with and without quantization are 2 \textmu s and 1.5 \textmu s, respectively.
    	\item SD\_Q, which combines SD with quantization, consumes 853 D flip-flops, which is 5.5x fewer than NFT\_Q and 2.7x  fewer than SST\_Q. Moreover, the synchronization errors of SD\_Q, NFT\_Q, SST\_Q, and SA\_Q are all below the minimum synchronization requirement of 8 \textmu s.
    \end{itemize}

    \textbf{Contributions:} We make the following contributions:
    \begin{itemize}
         \item We verify that optimizing cross-correlation synchronization in 5G backscatter cannot achieve both high accuracy and low resource consumption. Instead of using cross-correlation, we leverage symmetry to enable low-power and accurate PSS detection.
        \item We observe that the PSS envelope has a unique mirror symmetry. Based on this, we utilize differential techniques to convert multiplication into addition, reducing computational resources without losing accuracy.
        \item We have implemented our synchronization scheme on the prototype, and it meets the requirements of 5G backscatter in real scenarios. Empirical experiments confirm its practicality and effectiveness.
    \end{itemize}

\begin{figure}[!t]
\centering
\begin{minipage}[b]{\linewidth}
    \centering
        \subfloat[BER.]{
        \includegraphics[width=0.49\linewidth]{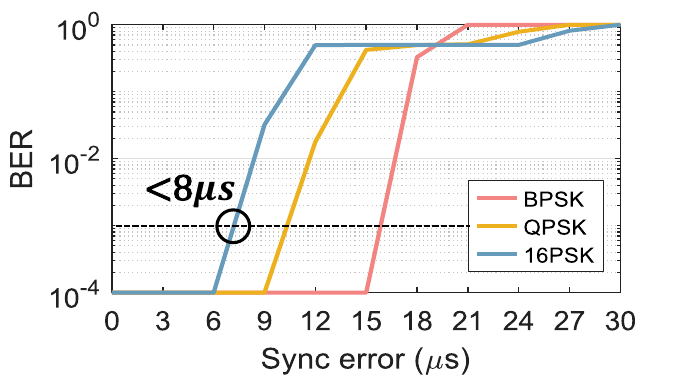}
        \label{error_ber}}
        \subfloat[Throughput.]{
        \includegraphics[width=0.47\linewidth]{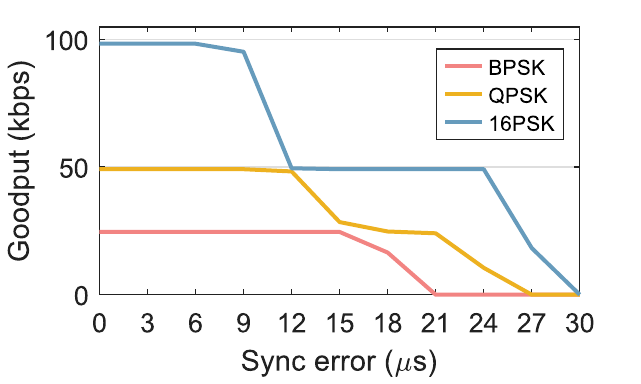}
        \label{error_throughput}}
    \caption{Impact of sync error on BER and throughput.}
    \label{error}
\end{minipage}
\end{figure}

\section{Motivation}

\subsection{Synchronization Requirements for 5G} \label{sec_requirement}

Synchronization is critical in both active and passive radios, as it aligns the receiver with incoming frames to locate symbol boundaries. The primary metric for synchronization is accuracy, which is fundamentally tied to the symbol duration of the excitation signal. WiFi OFDM symbols are 4 $\mu$s, LTE symbols are 66.67 $\mu$s, and 5G NR symbols vary with subcarrier spacing (e.g., 33.33 $\mu$s for 30 kHz spacing). Finer symbols require tighter synchronization to avoid ISI, explaining why 5G backscatter needs $\mu$s-level accuracy while LTE-based backscatter can tolerate tens of $\mu$s.
To quantify the required accuracy for 5G backscatter, we simulate a downlink using the 5G Toolbox (30 kHz SCS, 40 MHz bandwidth). We introduce timing offsets from 0 to 30 $\mu$s (step 3 $\mu$s) for BPSK, QPSK, and 16PSK under AWGN (SNR 15 dB) and measure BER over 1000 trials per offset. For reliable IoT applications, a BER $\leq 10^{-3}$ is widely accepted \cite{dunna2021syncscatter}. As shown in Fig.~\ref{error}, this requires sync error below 8 $\mu$s for the most sensitive modulation (16PSK). We adopt this stricter bound for robustness, setting the target to 8 $\mu$s. To illustrate its practical impact, consider low-power video streaming from wearable cameras: exceeding this error elevates BER above $10^{-3}$, rendering the stream unintelligible.

Another critical metric is resource consumption, as it directly impacts the power efficiency of synchronization—a paramount concern in backscatter systems. To assess feasibility, we select the Microsemi IGLOO nano AGLN250 FPGA as a representative ultra-low-power platform. According to its datasheet \cite{agln250}, the AGLN250 provides 6,144 VersaTiles, each configurable as a D flip-flop—fundamental elements for shift registers, counters, and arithmetic units. The resource usage of a synchronization algorithm is quantified by the number of D flip-flops required for its operations. Therefore, we set the resource budget for any practical 5G backscatter synchronization design to be within the 6,144 D flip-flop limit. Designs exceeding this budget cannot be implemented on this class of low-power devices without significant overhead. The AGLN250 is widely used in backscatter systems due to its low power \cite{yuan2023enabling,chi2020leveraging}. In the following sections, we evaluate methods against these two requirements: sync error $<8$ $\mu$s and resource consumption $<6,144$ D flip-flops.

\subsection{Why Previous Sync Methods Failed for 5G Backscatter?}
    \subsubsection{Active sync}
\begin{figure}[!t]
\centering
\begin{minipage}[b]{\linewidth}
    \centering
        \subfloat[CFO.]{
        \includegraphics[width=0.48\linewidth]{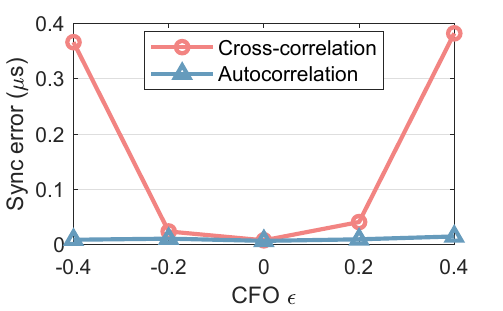}
        \label{cfo}}
        \subfloat[SNR.]{
        \includegraphics[width=0.48\linewidth]{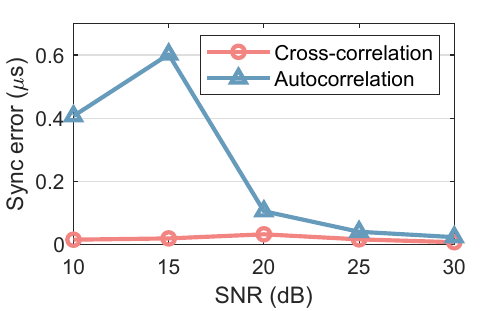}
        \label{snr}}
    \caption{Impact of CFO and SNR on correlation.}
    \label{cfo-snr}
\end{minipage}
\end{figure}

    Popular synchronization methods include autocorrelation and cross-correlation. 
    Active radios typically employ both autocorrelation and cross-correlation for timing and frequency synchronization. 
    Taking 5G as an example, due to the presence of Carrier Frequency Offset (CFO), the received signal can be represented as $y[n] = x[n] e^{j 2 \pi \epsilon n / N}$, where $x[n]$ is the transmitted signal and $N$ is the FFT size.
    5G User Equipment (UE) first applies autocorrelation using the repeated occurrence of cyclic prefixes. It multiplies the conjugate of the cyclic prefix with the corresponding tail to estimate and eliminate the CFO: $\hat{\epsilon} = \frac{1}{2\pi} \arg \left\{ \sum_{n=-N_G}^{-1} y^*[n] y[n+N] \right\}$.
    Moreover, this step achieves coarse synchronization. 
    For fine synchronization, after eliminating the CFO, UE cross-correlates it with three pre-stored possible PSSs $p[n]$, indexed by $i$, where $C(\theta, i) = \frac{|\sum_{n=0}^{N-1} y[\theta+n] p^*[n]|}{\sum_{n=0}^{N-1} |y[\theta+n]|^2}$.
    The synchronization process benefits from autocorrelation's robustness against frequency offsets and cross-correlation's strong noise immunity capabilities.
    We evaluate the impact of CFO and SNR on cross-correlation and autocorrelation, with a subcarrier spacing of 30 kHz. Here, $\epsilon$ is set to 0.2, and the CFO is 6 kHz.
    As depicted in Fig. \ref{cfo}, we observe that autocorrelation exhibits greater tolerance to frequency offset than cross-correlation. Outside the range of $\epsilon$ from -0.2 to 0.2, the synchronization error of autocorrelation is 0.4 \textmu s lower than cross-correlation. As shown in Fig. \ref{snr}, it is evident that cross-correlation shows superior noise resistance compared to autocorrelation. When SNR is below 20 dB, the synchronization error of cross-correlation is 0.6 \textmu s lower than autocorrelation.
    However, active sync is prohibitive for 5G tags because it requires a high-power analog-to-digital converter (ADC) and voltage-controlled oscillator (VCO), as well as performing computation-intensive FFT.

\begin{figure}
	\centering
	\includegraphics[width=0.95\linewidth]{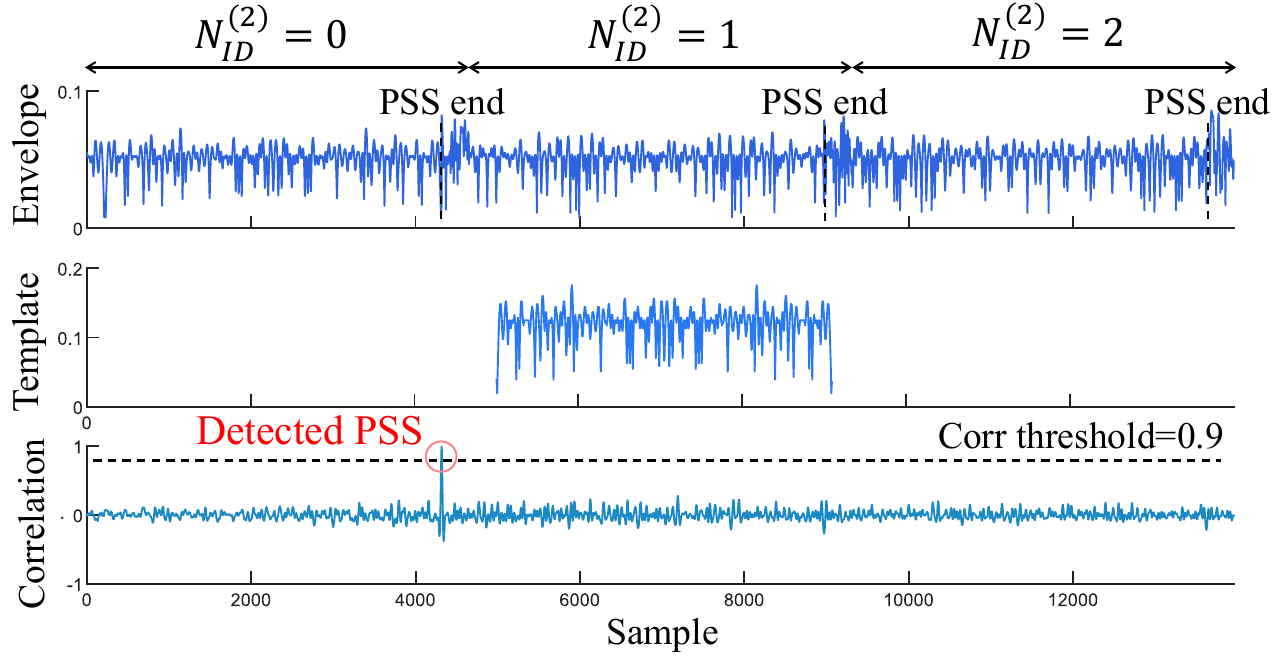}
	\caption{Traditional sync using one template where \( N_{ID}^{(2)} = 1 \) for 5G backscatter. The tag can detect PSS only when the incoming 5G signal's $N_{ID}^{(2)}$ is 0; otherwise, synchronization fails due to correlation coefficients far below the threshold.}
	\label{one_template}
\end{figure}

    \subsubsection{Passive sync}\label{sec_challenge}
    Given the resource constraints of backscatter radio and its much lower synchronization requirements than active radios, the input to tags is an envelope rather than IQ data. This means that multiplying the conjugate of the cyclic prefix with the same tail for CFO estimation and cancellation is not feasible, and autocorrelation loses its ability to combat frequency offset. As a result, the combination of autocorrelation and cross-correlation fails to achieve the expected effectiveness. Therefore, traditional passive synchronization methods such as PLoRa \cite{peng2018plora} and Multiscatter \cite{gong2020multiprotocol} often resort to cross-correlation for timing. However, using one template from traditional synchronization methods (e.g., PLoRa and Multiscatter) for 5G backscatter can lead to synchronization failure. For example, Multiscatter achieves synchronization with the excitation signal using template matching and quantization techniques. Fig. \ref{one_template} illustrates the effect of Multiscatter applied to 5G backscatter, where the template chosen is PSS with cell ID sector \( N_{ID}^{(2)} = 0 \). \( N_{ID}^{(2)}\) determines the PSS sequence, and \( N_{ID}^{(2)}\) can take one of three values: 0, 1, or 2. We observe that when the incoming 5G signal contains \( N_{ID}^{(2)} = 1 \) or \( 2 \), this method fails to detect the PSS. The reason for this is that 5G PSS consists of three different sequences, requiring three distinct templates. Similarly, the existing LTE backscatter system SyncLTE \cite{feng2023heartbeating} also adopts a one-template approach. Unlike MultiScatter, SyncLTE exploits the periodicity of LTE signals and designs a discontinuous template to further enhance synchronization accuracy. However, such one-template methods are not applicable to all 5G signals and are limited to signals with specific cell IDs. In addition, LScatter \cite{chi2020leveraging}, an advanced LTE backscatter system, achieves synchronization by detecting signal strength variations at the rising edge of PSS. Although this approach is simple to implement, it suffers from low synchronization accuracy due to the complex structure of 5G frames, which consist of various intertwined time-frequency components, making it difficult to distinguish the PSS from other signals.
    Moreover, methods designed for intermittent signals, such as SyncScatter \cite{dunna2021syncscatter} (which uses rising-edge detection for WiFi backscatter), are ill-suited for 5G. WiFi signals are bursty with clear idle periods, allowing simple energy detection to locate packet boundaries. In contrast, 5G NR signals are continuously transmitted with no idle gaps; the frame structure multiplexes various channels (PSS, SSS, PBCH, data) in time and frequency. Rising-edge detection on the envelope often mistakes other symbols (e.g., PBCH) for the PSS, leading to frequent false synchronization. This issue is exacerbated in dense urban environments where signal variations are high. Therefore, a more robust and resource-efficient synchronization method tailored to 5G's continuous and multi-cell nature is essential.

\begin{figure}[!t]
\centering
\begin{minipage}[b]{\linewidth}
    \centering
        \subfloat[Without symmetry.]{
        \includegraphics[width=0.5\linewidth]{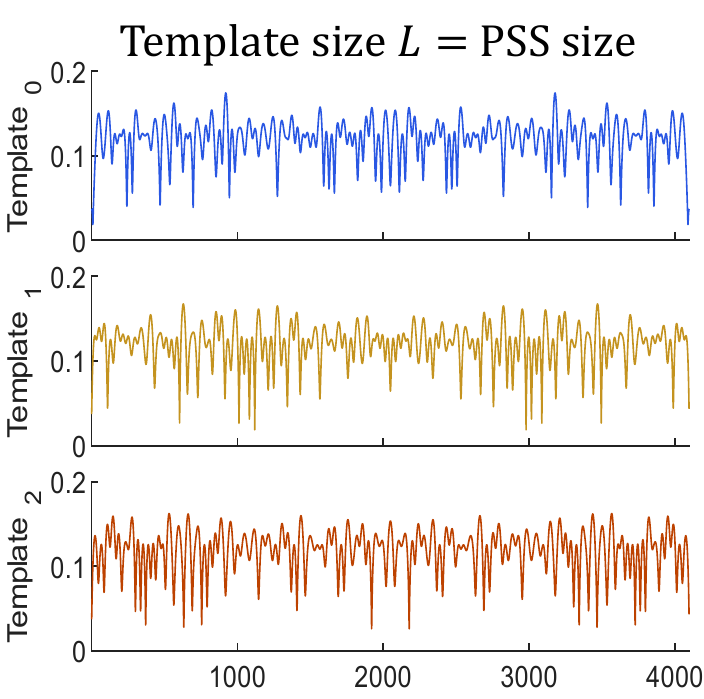}
        \label{wo_symmetry}}
        \subfloat[With symmetry.]{
        \includegraphics[width=0.5\linewidth]{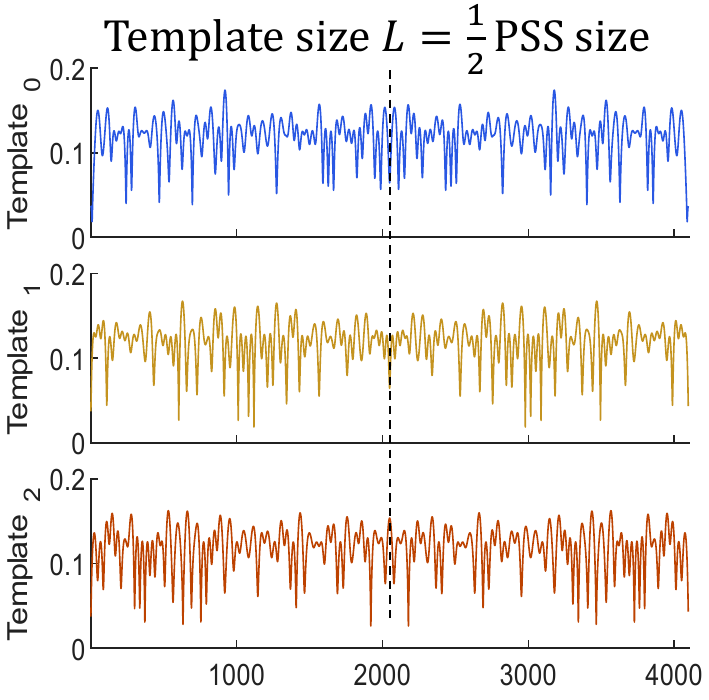}
        \label{mirror_symmetry}}
    \caption{5G PSS envelope exhibits mirror symmetry. We can halve the template size by leveraging its mirror symmetry.}
    \label{symmetry}
\end{minipage}
\end{figure}

\section{Synchronization Design}
We first verify that cross-correlation cannot be used for 5G backscatter. Here, we propose multi-template NFT to address the limitations of one-template sync and optimize resources with SST. 
Furthermore, we design low-power sync for 5G backscatter based on symmetry. 
Initially, we present SA, which still consumes more resources than the tag's capacity. 
Therefore, we have devised SD to convert multiplication into addition, fulfilling the requirements for low-power sync.
\subsection{Cross-correlation-based Sync for 5G Backscatter}
\subsubsection{NR Fine Timing Sync}
    Based on 5G NR fine timing (NFT), a straightforward approach involves cross-correlating with all templates, using peak correlations to determine synchronization points. 
    Despite achieving the desired synchronization accuracy, this method requires substantial computational resources. 
    Passive cross-correlation is expressed as \begin{small}\begin{align*} r_i(t) = \frac{\sum_{n=0}^{L-1} (S_{n+t} - \bar{S})(P(i)_n - \bar{P}(i))}{\sqrt{\left( \sum_{n=0}^{L-1} (S_{n+t} - \bar{S})^2 \right) \left( \sum_{n=0}^{L-1} (P(i)_n - \bar{P}(i))^2 \right)}} \tag{1} \label{f_1}\end{align*}\end{small}
    where $S$ is the received 5G envelope, $t$ is the time delay, $L$ is the template size (equal to PSS size $\rho$), $P(i)$ represents the $i$-th PSS (i $\in \{0, 1, 2\}$), and $\bar{S}$ and $\bar{P}$ are their respective means.
    According to Eq.(\ref{f_1}), NFT cross-correlation requires $3(3\rho+1)$ multipliers and $3(5\rho-3)$ adders, where $L=\rho$.
    Next, we consider the feasibility of implementing these computations on a typical low-power AGLN250 FPGA. Here, with a 1 MHz sampling rate and a template size of 36 bits per symbol length, a $12\times12$ multiplier and adder, respectively, require 456 and 25 D flip-flops.
    Consequently, NFT requires 162,387 D flip-flops, far exceeding the total of 6,144 D flip-flops available on the AGLN250.

    To achieve this goal, we first optimize its resource utilization along the path of cross-correlation-based synchronization, exploring feasible solutions.
    Hence, an immediate concern arises: how can we reduce the computational complexity of cross-correlation-based synchronization? First, we analyze the factors influencing its computational complexity. According to Eq.(\ref{f_1}), the computational load of one cross-correlation execution $C(r(t)) = 3(9L - 1) = O(L)$. Hence, the total computational load $\Psi = M \times C(r(t)) = O(ML)$, where $L$ is the template size, and $M$ is the number of cross-correlation executions. It is evident that the total computational load scales linearly with $ML$. Therefore, reducing $L$ and $M$ is essential to minimize the overall computational requirement.

\begin{figure}[!t]
	\centering
	\includegraphics[width=0.94\linewidth]{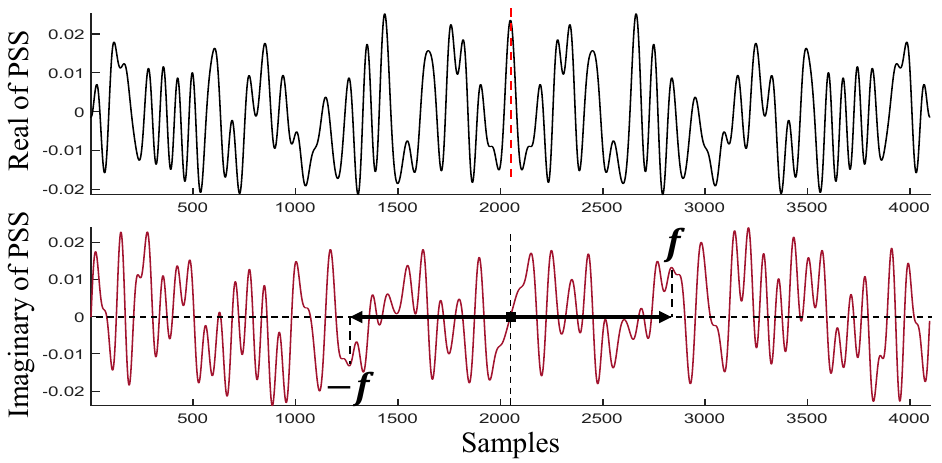}
	\caption{The real and imaginary parts of PSS. The real and imaginary parts of the PSS exhibit even and odd symmetry, respectively.}
	\label{pss}
\end{figure}

\subsubsection{Symmetry-based Semi-Template Synchronization} \label{sec_sst}
An intriguing observation in Fig.~\ref{symmetry} is that the PSS envelope exhibits mirror symmetry, which allows us to reduce the template size $L$ to half the size of the PSS. But why does this symmetry exist? Such a symmetry arises from the conjugate symmetry of NR PSS. Next, we prove its existence. The frequency domain NR PSS is a Binary Phase Shift Keying (BPSK) M-sequence with 127 real values of $+1$ and $-1$, denoted as $\{X(k) \mid k = 0, 1, \ldots, 126\}$. Upon Inverse Discrete Fourier Transform (IDFT), we obtain the PSS time-domain waveform, denoted as $\{P(n) \mid n = 0, 1, \ldots, N-1\}$, where $N$ is the IDFT size. Hence,
$
P(n) = \text{IDFT}[X(k)] = \frac{1}{N}\sum_{k=0}^{N-1} X(k) e^{j2\pi kn/N}.
$
The complex conjugate of $P(N-n)$ is
\[
\begin{small}
    \begin{aligned}
P^*(N-n) &= \left( \frac{1}{N}\sum_{k=0}^{N-1} X(k) e^{j2\pi k(N-n)/N} \right)^* \\
         &= \frac{1}{N}\sum_{k=0}^{N-1} X^*(k) e^{-j2\pi k(N-n)/N} \\
         &= \frac{1}{N}\sum_{k=0}^{N-1} X(k) e^{j2\pi k n / N} = P(n),
\end{aligned}
\end{small}
\]
where we used the fact that $X(k)$ is real ($X^*(k)=X(k)$) and $e^{-j2\pi k}=1$. Therefore, the NR PSS in the time domain exhibits complex conjugate central symmetry. Fig.~\ref{pss} illustrates the real and imaginary parts of the PSS, where we observe that its real part is even symmetric and its imaginary part is odd symmetric, further confirming its conjugate central symmetry.

This conjugate symmetry directly implies mirror symmetry in the envelope domain. Specifically, taking the modulus on both sides of $P(n) = P^*(N-n)$ yields $|P(n)| = |P(N-n)|$, which indicates that the envelope of the PSS waveform is symmetric about its center. This property is inherent to all NR PSS sequences, as it stems from the frequency-domain PSS being a real-valued BPSK M-sequence. After IDFT, the time-domain signal inherits conjugate symmetry, and the envelope operation further converts it into mirror symmetry.

\begin{figure}[!t]
	\centering
	\includegraphics[width=0.98\linewidth]{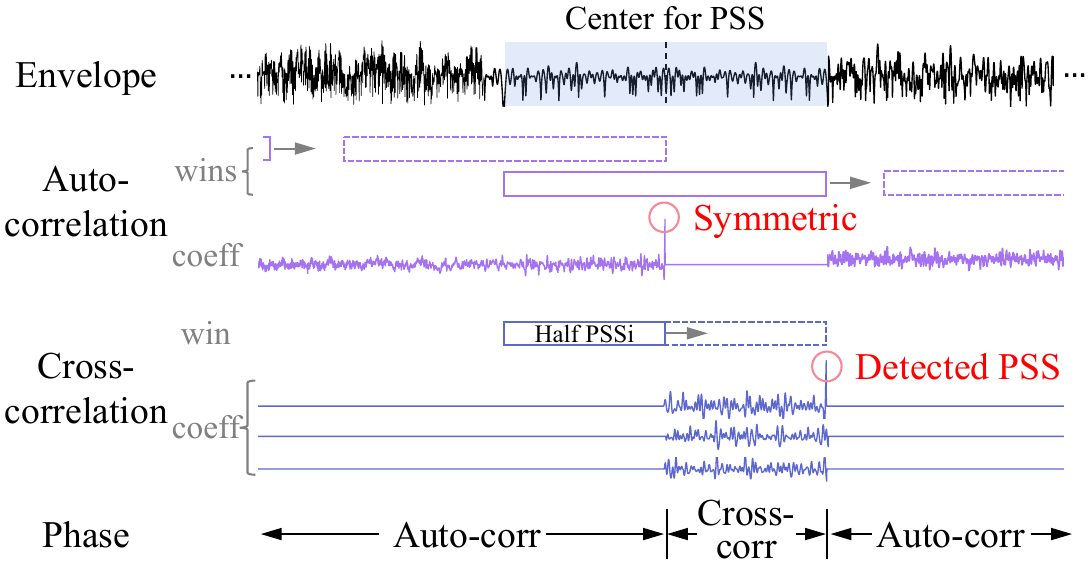}
	\caption{Symmetry-based semi-template synchronization. SST utilizes the mirror symmetry of the PSS envelope: it first uses autocorrelation to find the symmetric position, then cross-correlates with half of the three PSSs to locate the PSS.}
	\label{sst}
\end{figure}

    Based on this observation, we propose symmetry-based semi-template sync (SST), which utilizes the mirror symmetry of the PSS envelope to reduce the number of cross-correlation executions $M$ and the template size $L$, thereby lowering the computational complexity of correlation-based synchronization. As shown in Fig. \ref{sst}, SST first applies autocorrelation to determine if the received 5G envelope is mirror symmetric. Once symmetry is confirmed, SST performs cross-correlation between the 5G envelope and half of all templates, using the peak correlation values to detect the PSS positions. It is evident that the time consumption for cross-correlations during synchronization is minimal since the minimum period of PSS appearances is 5 ms, and the template size is only half the size of the PSS.
    With the template size halved, the computational resources required for cross-correlation amount to $3(3\rho/2+1)$ multipliers and $3(5\rho/2-3)$ adders, totaling 81,765 D flip-flops. However, SST still cannot work efficiently on low-power AGLN250 due to the high resource consumption associated with cross-correlation using three templates.
    In fact, the computational resources required for SST's cross-correlation are directly proportional to the template size. Ways to reduce the template size include (a) reducing the length of the template itself and (b) decreasing the sampling rate.
    Hence, in addition to reducing the length of the template itself, we can further decrease the sampling rate to reduce the template size. However, this still cannot achieve a trade-off between synchronization accuracy and resource consumption.
    Section \ref{sec_eva} evaluates the synchronization error and resource usage across sampling rates, illustrating that SST struggles to achieve both low resource consumption and high accuracy.

\subsection{Ultra-low Power Synchronization for 5G Backscatter} 
\begin{figure*}[!t]
\centering
\begin{minipage}[b]{\linewidth}
    \centering
        \subfloat[Symbol \#0.]{
        \includegraphics[width=0.29\linewidth]{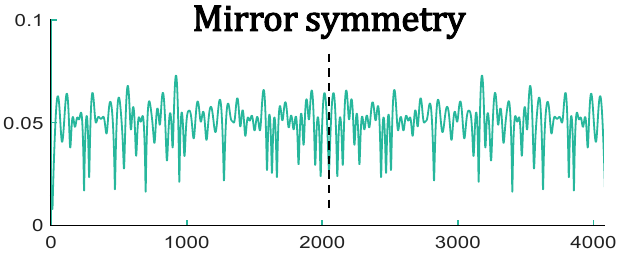}}
        \hfill
        \subfloat[Symbol \#1 and \#3.]{
        \includegraphics[width=0.29\linewidth]{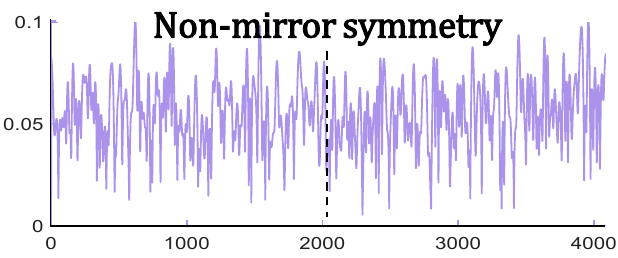}}
        \hfill
        \subfloat[Symbol \#2.]{
        \includegraphics[width=0.29\linewidth]{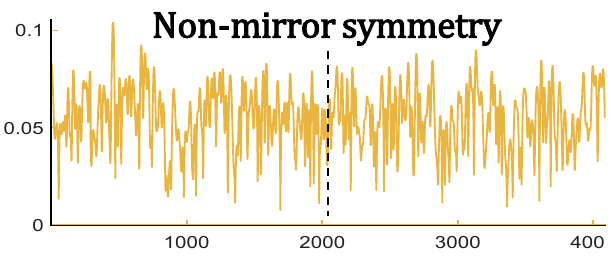}}
    \caption{The NR PSS envelope exhibits a unique mirror symmetry. Symbol \#0, where PSS is located, is the only one that is mirror symmetric within the envelopes of all symbols in SSB.}
    \label{unique_symmetry}
\end{minipage}
\end{figure*}

    Another critical challenge is how to design ultra-low power synchronization for 5G backscatter. 
    A key observation is that the mirror symmetry of the PSS envelope is not only inherent but also unique among 5G signals. 5G utilizes Synchronization Signal Block (SSB) for downlink synchronization, where symbol \#0 exclusively contains PSS, symbols \#1 and \#3 are composed of Physical Broadcast Channel (PBCH) carrying variable MIB due to the SFN field, and symbol \#2 consists of PBCH and Secondary Synchronization Signal (SSS), also variable. Fig.~\ref{unique_symmetry} illustrates the envelopes of different SSB symbols, revealing that only symbol \#0 exhibits mirror symmetry, confirming the uniqueness of the PSS envelope symmetry. This uniqueness extends beyond the SSB: data symbols in 5G NR carry user-specific information that is scrambled and modulated with higher-order schemes (e.g., QPSK, 16QAM) and lack any structural regularity like the conjugate symmetry of PSS. Consequently, their envelopes do not exhibit mirror symmetry, making the mirror symmetry a unique fingerprint of the PSS that enables reliable detection without false alarms from other symbols. This exclusivity is a key enabler for our low-power synchronization method.
    Therefore, by leveraging this feature, we can detect PSS by identifying the position of its mirror symmetry. Although this symmetry-based approach lacks the noise resilience of cross-correlation—making it challenging under low SNR—it offers significant potential for low-power synchronization. Importantly, the envelope symmetry is robust to CFO because the envelope operation discards phase information, but it remains susceptible to additive noise that distorts the envelope shape and increases synchronization error under low SNR conditions.

    \subsubsection{Symmetric Autocorrelation-based Sync}
    A simple method is to use autocorrelation within SST to find the symmetric position and locate the center of PSS, which we call symmetric autocorrelation (SA)-based sync. 
    Autocorrelation consumes computational resources equivalent to one-third of the resources required for half of PSS cross-correlation, as it correlates only with the symmetric position without requiring three templates. 
    The computational resources for SA's autocorrelation depend on the sliding window size, which is fixed at the size of one PSS. The only way to reduce the window size is by decreasing the sampling rate.
    Consequently, when the sampling rate is 1 MHz, and the window size corresponds to a 36-bit PSS size, it consumes 55 multipliers, 87 adders, and 27,255 D flip-flops, which still exceeds the resource capacity of AGLN250. 
    However, SA requires a minimum sampling rate greater than 1 MHz to meet the desired accuracy requirements, indicating that it consumes even more resources.
    The primary reason is that multipliers consume far more resources than adders, about 18 times more. 
It is worth noting that methods like NFT, SST, and SA are designed for scenarios with abundant computational resources—such as active UEs—where multiplier power consumption is acceptable. However, in the target scenario of a 5G backscatter tag, e.g., a battery-free vibration sensor for structural health monitoring that operates on harvested energy (tens of microwatts), such resource-heavy approaches become infeasible. Their high hardware resource demands would overwhelm the tag's energy budget. 

\begin{figure}[!t]
	\centering
	\includegraphics[width=0.95\linewidth]{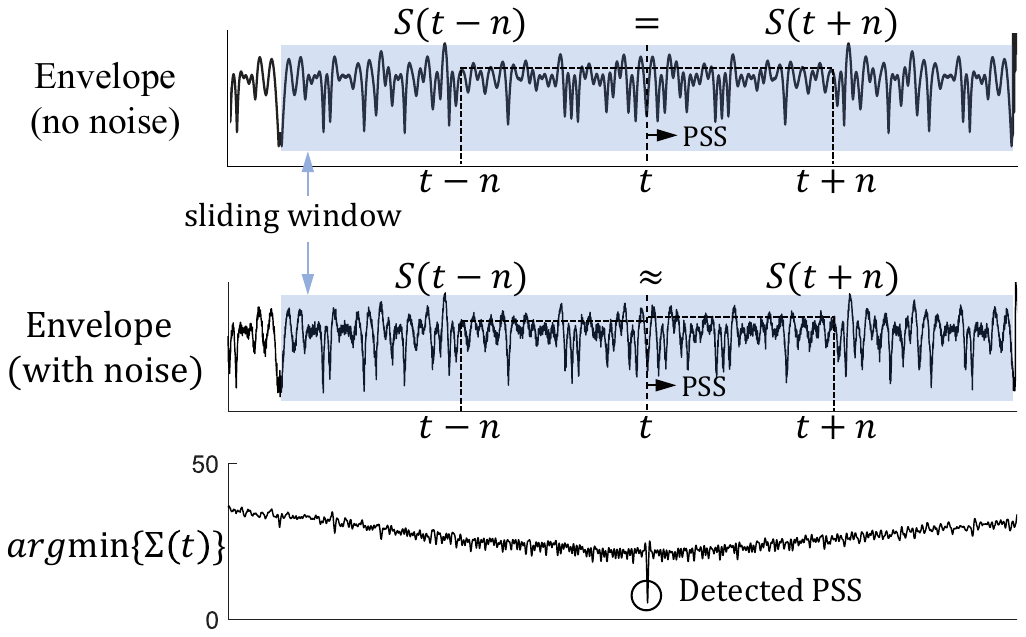}
	\caption{Symmetric differential-based sync. SD computes $\Sigma(t)$ using differential summation within a sliding window, and the PSS is detected at the minimum of $\Sigma(t)$.}
	\label{sd}
\end{figure}

\subsubsection{Symmetric Differential-based Sync }
To address this issue, a key idea is to convert multipliers into adders to reduce computational resource consumption. Section~\ref{sec_sst} proves that the PSS exhibits complex conjugate symmetry, i.e., $P(n)=P^{*}(N-n)$, which directly implies $|P(n)|=|P(N-n)|$ and thus $|P(n)|-|P(N-n)|=0$. Exploiting this property, we propose the Symmetric Differential (SD) synchronization method. As shown in Fig.~\ref{sd}, SD computes a symmetry metric $\Sigma(t)$ within a sliding window of length $\rho$ (one PSS duration) by summing the absolute differences between symmetric envelope samples: \begin{small}
    \begin{align*}\Sigma(t)=\sum_{n=1}^{\rho/2}\left|S(t+n)-S(t-n)\right| \tag{2} \label{f_2}\end{align*}
\end{small}
where $t$ denotes the window's center moment. In an ideal noise-free scenario, $\Sigma(t)=0$ exactly at the PSS center. Under practical noise conditions, the envelope is perturbed, but the minimum of $\Sigma(t)$ over the sliding window remains distinct as long as the SNR is above a certain threshold. Thus, PSS detection reduces to finding $\arg\min\{\Sigma(t)\}$.

It is important to distinguish our SD approach from prior symmetry-based methods, such as the AHC and CSC detectors in \cite{zhang2012low}. Although \cite{zhang2012low} leverages the central symmetry of LTE PSS to reduce computational complexity in active UE synchronization, it still relies on complex multiplications (e.g., reducing from N to N/2+1 multiplications per PSS) and targets IQ samples. In contrast, our SD method operates on envelope samples for passive 5G backscatter tags and completely eliminates multiplications by replacing them with simple differential additions. This multiplier-to-adder conversion is enabled by the unique mirror symmetry of the 5G PSS envelope and the envelope equality at symmetric points $|P(n)|=|P(N-n)|$. According to Eq.(\ref{f_2}), SD consumes zero multipliers and $\rho-1$ adders, requiring only 875 D flip-flops—well below the 6,144 D flip-flop capacity of the ultra-low-power AGLN250 FPGA. To the best of our knowledge, this is the first multiplier-free PSS detection scheme for cellular backscatter, enabling ultra-low-power synchronization on resource-constrained tags. However, its reliance on envelope symmetry also makes it vulnerable to noise and multipath-induced distortions: under low SNR, the synchronization error increases compared to cross-correlation-based methods such as NFT and SST, which offer stronger resilience. To balance power efficiency and accuracy across different channel conditions, we adopt a hybrid strategy that adaptively combines SD and SST\_Q (SST with quantization). Specifically, the tag estimates the current channel quality and selects SD in favorable conditions for low-power operation, while switching to SST\_Q when the channel degrades. This adaptive approach requires no pre-deployment calibration and maintains robust synchronization under both noise and multipath fading.

    


Real-time detection is achieved by streaming envelope samples through a shift register of length $\rho$, computing $\Sigma(t)$ every clock cycle using parallel adders and absolute-value circuits. Since the operations involve only additions and comparisons (no multiplications), the design can operate at the sampling rate on ultra-low-power FPGAs, with resource consumption far below the available budget. This hardware-friendly architecture makes SD suitable for continuous synchronization in 5G backscatter tags.
It is worth noting that our SD method is platform-agnostic. Its resource consumption, determined by the number of adders and comparators, scales linearly with the sampling rate and can be mapped to any FPGA with sufficient logic elements. The choice of AGLN250 represents a typical ultra-low-power platform, but the relative advantages of SD—low resource usage and high accuracy—hold across different hardware implementations.

    SD meets the requirement for low computational resource consumption, but its overall computational load is high because it increases linearly with the number of samples, equal to $N(\rho-1)$, where $N$ is the number of samples. 
    For example, when tags sample 5G signals at 5 MHz and the minimum PSS period is 5 ms, if the total duration exactly covers 10 PSS periods, the total sample count is $N=5 \times 5 \times 1000 \times 10=250,000$, and $\rho=167$, so the total computational load for SD amounts to $4.15 \times 10^{7}$, which is quite substantial.
    In addition, the overall computational loads of NFT, SST, and SA also increase linearly with the number of samples, and they are significantly higher than that of SD.
    To address this, we propose SD+, which still utilizes SD differentials for initial PSS detection. 
    Once the first PSS is identified, SD+ leverages the periodicity of PSS by adding integer multiples of the PSS cycle to determine subsequent PSS instances. 
    SD+ significantly reduces the overall computational load by only employing a differential technique for initial PSS detection, scaling down to approximately $1/N\sim25,000/N$ of SD's load.
    Note that methods such as NFT, SST, and SA, even when they utilize periodicity to reduce the overall computational load, do not alter their resource consumption; they still exceed the resource capacity of the tag.

\begin{figure}[!t]
\centering
\begin{minipage}[b]{0.19\linewidth}
    \centering
    \subfloat[Tag]{
        \includegraphics[width=\linewidth]{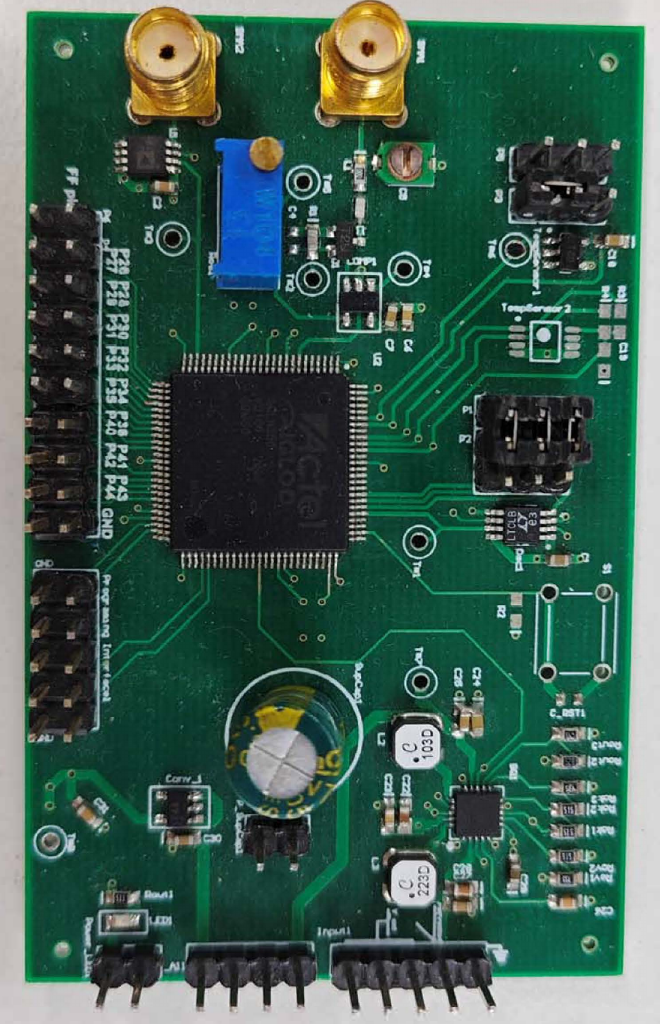}
        \label{prototype}}
\end{minipage}
\hfill
\begin{minipage}[b]{0.16\linewidth}
    \centering
    \subfloat[gNodeB]{
        \includegraphics[width=\linewidth]{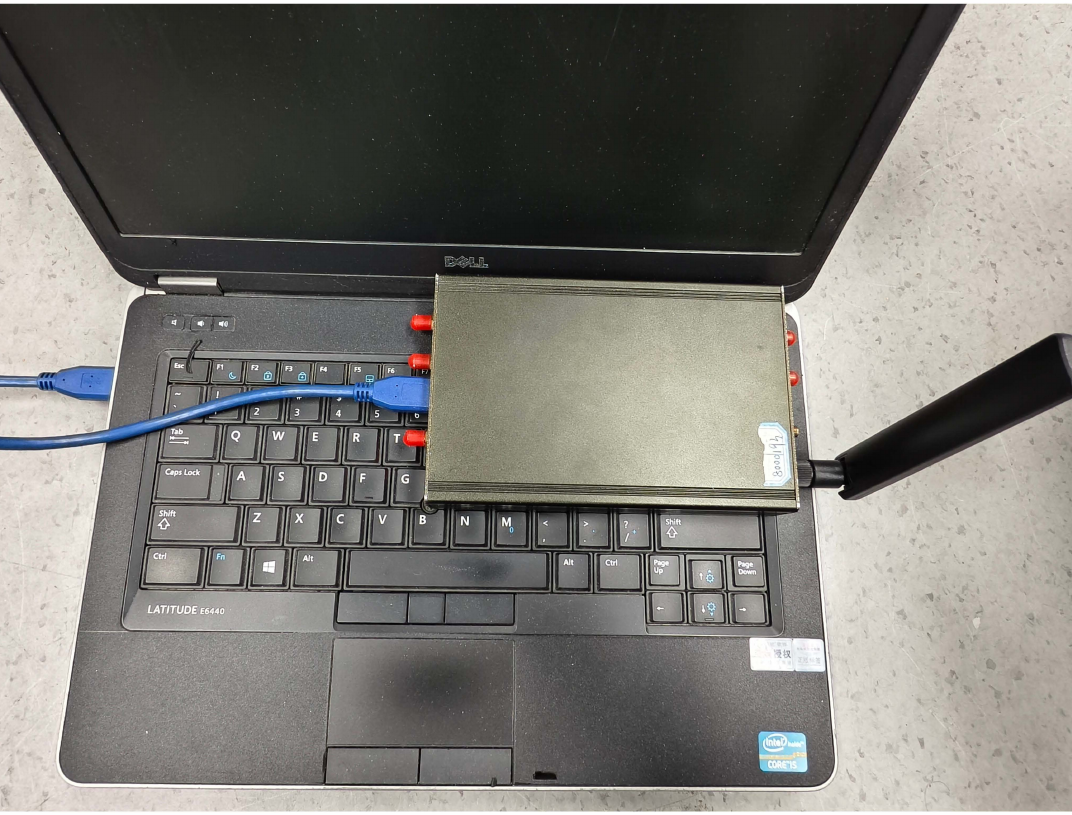}
        \label{gnodeb}}
    \vspace{-0.9em}
    
    \subfloat[UE]{
        \includegraphics[width=\linewidth]{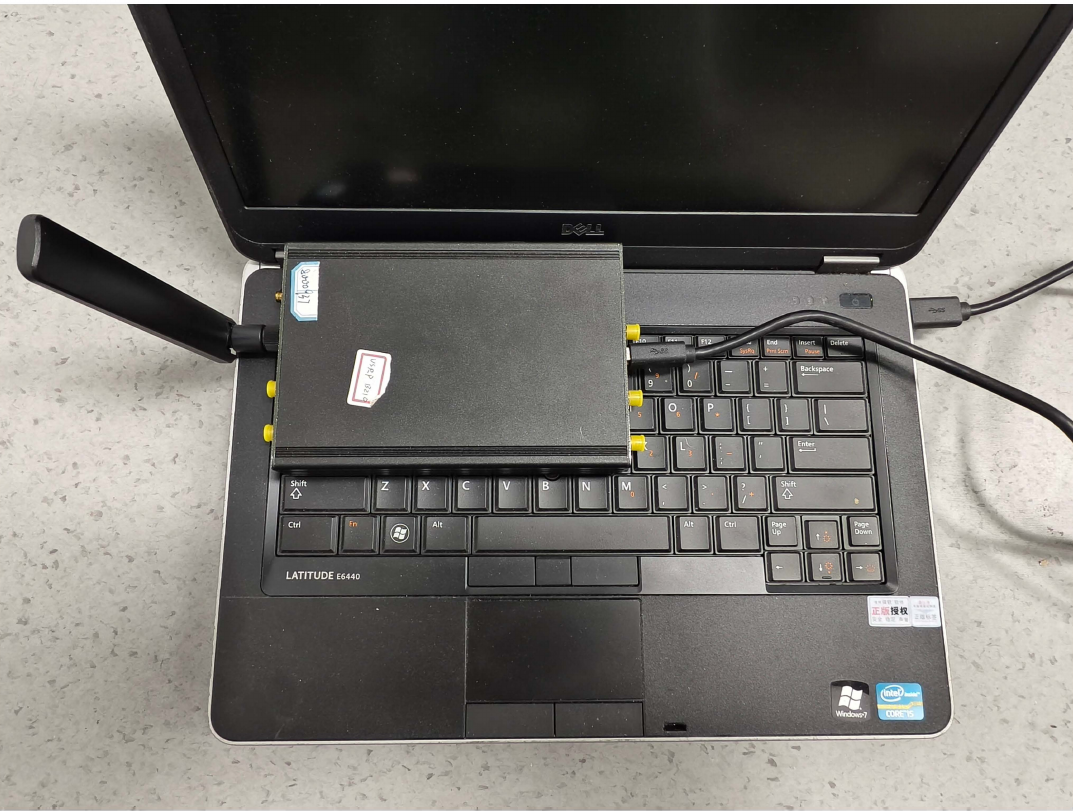}
        \label{ue}}
\end{minipage}
\hfill
\begin{minipage}[b]{0.62\linewidth}
    \centering
    \subfloat[Experimental deployment]{
        \includegraphics[width=\linewidth]{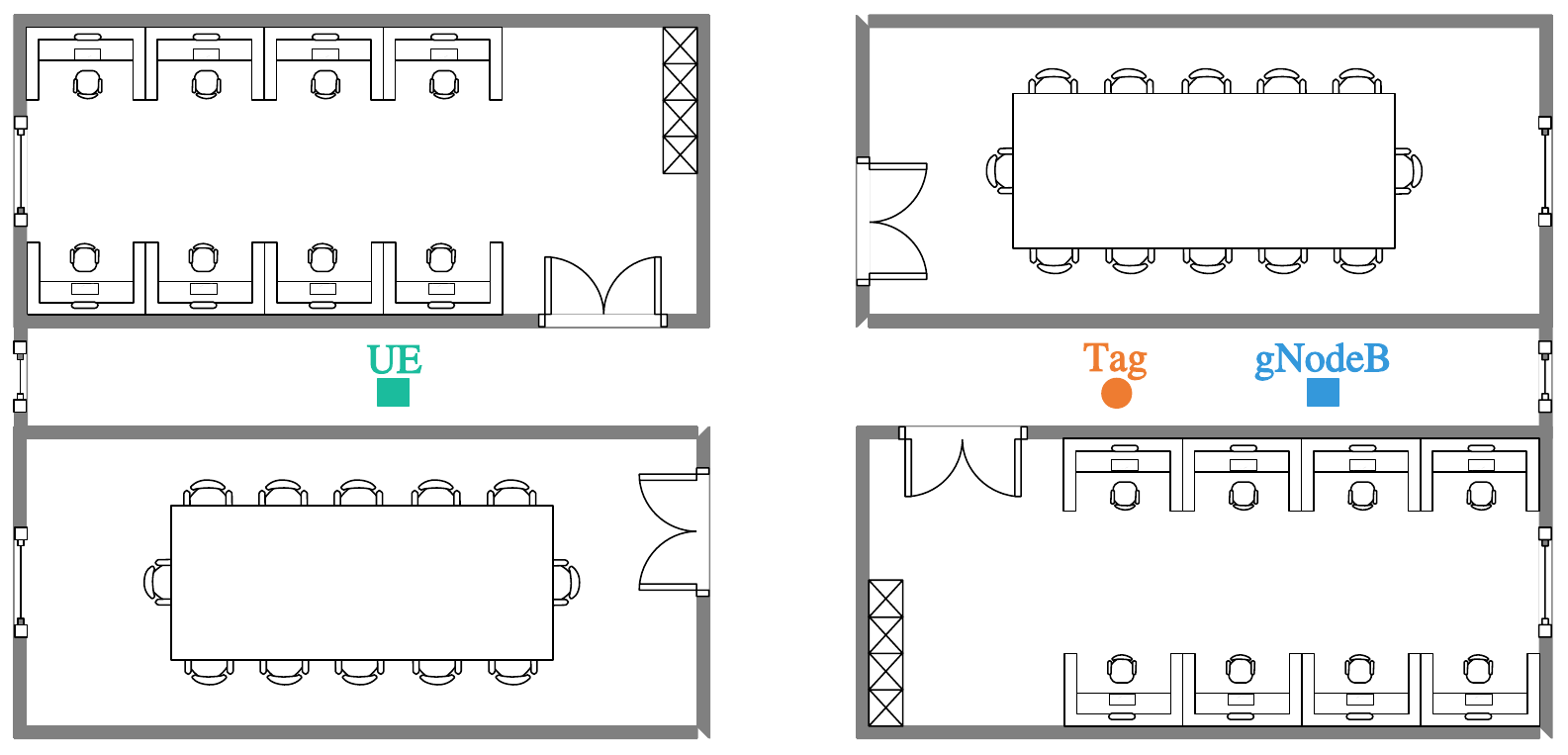}
        \label{deployment}}
\end{minipage}

\caption{System prototype and experimental deployment.}
\label{system}
\end{figure}
\section{Implementation}

In our real-world experiments, we first implement the SD method for a low-power tag prototype and utilize quantization techniques from Multiscatter \cite{gong2020multiprotocol} to achieve quantized versions of all sync methods on low-power FPGAs.
Quantization significantly reduces power consumption compared to non-quantization methods using high-energy ADCs, employing comparators instead. 
This adaptation enables sync methods like NFT, SST, and SA, which are impractical for tags due to their resource demands, to operate effectively by converting multiplication operations into additions.
Key differences between quantization and non-quantization include:
(a) Sampling depth: Non-quantization typically uses a 12-bit ADC, while quantization utilizes a 1-bit comparator.
(b) Power consumption: ADCs consume considerably more power than comparators, and quantization simplifies computational complexity.
Given the high resource consumption of NFT, SST, and SA, which precludes their implementation on low-power FPGAs, we supplement our experimental efforts with MATLAB simulations to assess these sync methods for backscatter.


\noindent{\bf Practical Implementation:}
    (1) Verification Prototype: Fig.~\ref{prototype} shows our tag verification prototype that consists of the following components: synchronization algorithms implemented on a low-power Microsemi Igloo Nano AGLN250 FPGA. The RF front-end includes a customized passive envelope detector and either an NCS2200 voltage comparator or LTC2366 ADC.
    (2) IC Prototype: We simulate an IC prototype using Cadence IC6.17 Virtuoso software and TSMC 0.18 \textmu m CMOS process design kits tailored for low-power design. 
    This prototype includes a detector, an oscillator, and a digital core with synchronization capabilities. Additionally, we run OpenAirInterface (OAI) \cite{oai} with USRPs B210 \cite{USRP} as the 5G exciter and UE as shown in Figs.~\ref{gnodeb} and~\ref{ue}. Moreover, Fig.~\ref{deployment} shows our experimental deployment.
    Note that we employ an RF front-end with a detector and ADC for SD implementation and another with a detector and comparator for quantized versions of all synchronization methods.
    Multiplication for correlation-based synchronization is quantized to addition, enabling functionality on low-power FPGAs.

\noindent{\bf MATLAB Simulation:}
    Firstly, we use MATLAB APP \cite{matlabAPP} and Toolbox \cite{matlabToolBox} to obtain 5G downlink waveforms and PSS. Then, we extract their envelopes and digitize them using ADC to obtain digital 5G envelopes and templates. Finally, we implement the following five synchronization algorithms:
    (1) NFT: Cross-correlation with three PSSs for PSS detection.
    (2) SST: Autocorrelation for symmetry determination and three half-PSS cross-correlations for PSS detection.
    (3) SA: Autocorrelation for PSS detection.
    (4) SD: Differential for PSS detection.
    (5) SD+: Differential for initial PSS detection and PSS periodicity for subsequent PSS detection.
    Note that SD+ outperforms SD in total computational load while maintaining identical synchronization accuracy and resource consumption.

\noindent{\bf Competitors:}
    (1) In real-world experiments, we first compare SD with and without quantization in terms of accuracy and power consumption. We then evaluate their quantized versions (NFT\_Q, SST\_Q, SA\_Q, SD\_Q). Note that SD\_Q and SA\_Q correspond to XOR accumulation and its reverse, yielding similar performance. We also compare SD\_Q with existing backscatter synchronization methods: SyncLTE \cite{feng2023heartbeating} and Multiscatter \cite{gong2020multiprotocol} (both using one-template matching with PSS of cell ID $N_{ID}^{(2)}=0$), and LScatter \cite{chi2020leveraging} (based on PSS rising edge detection). NFT\_Q is the multi-template version of Multiscatter. In MATLAB simulations, we compare SD with NFT, SST, and SA in terms of accuracy, delay, resource consumption, and total computational load.
    
     (2) Metrics: (a) Synchronization error: defined as the time jitter between the detected PSS and true PSS. (b) Synchronization delay: defined as the time taken to detect one instance of PSS. (c) Total computational load: defined as the total number of operations performed during the runtime.
     (d) Synchronization success rate: defined as the proportion of total synchronization attempts in which the synchronization error is less than 8 \textmu s. An attempt is classified as successful if the error falls below this threshold; otherwise, it is considered a failure.
 
\section{Evaluation}
\subsection{End-to-End Performance}
\subsubsection{Comparison of Sync Methods with quantization}
\begin{figure} [t]
	\centering
        \subfloat[Sync error.]{\includegraphics[width=0.45\linewidth]{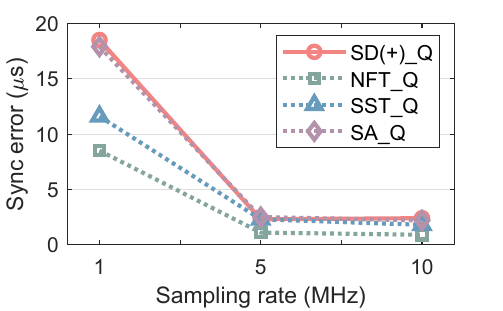}
			\label{eva_error_sr_Q}}
		\hfil
		 \subfloat[Resource consumption.]{\includegraphics[width=0.45\linewidth]{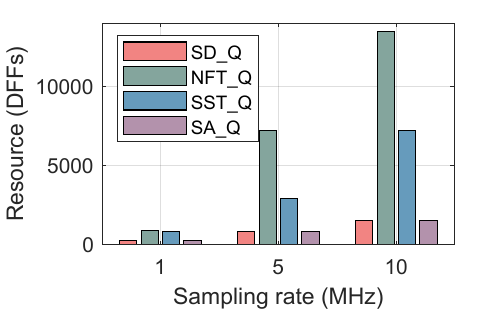}
			\label{eva_resource}}
	\caption{Sync performance with quantization.}
 \hfil
\end{figure}

\ 
\newline
\indent \noindent{\bf Synchronization Accuracy:}
    We evaluate the synchronization errors of various methods combined with quantization across different sampling rates. 
    Section \ref{sec_requirement} mentions that to achieve a BER below $10^{-3}$, the synchronization error should be less than 8 \textmu s. 
    As shown in Fig. \ref{eva_error_sr_Q}, we observe that all synchronization methods require a minimum sampling rate of 5 MHz to meet this criterion.
    The synchronization errors for SD(+)\_Q, NFT\_Q, SST\_Q, and SA\_Q at 5 MHz sampling rate are 2.3 \textmu s, 1.1 \textmu s, 2.3 \textmu s, and 2.5 \textmu s, respectively, all very close to each other. 
    Additionally, we note that the synchronization performance of SD(+)\_Q and SA\_Q is remarkably similar. 
    This similarity arises because the implementation of SD(+)\_Q involves accumulating XOR results of symmetrically positioned pairs of bits (0 or 1) and finding the minimum, while SA\_Q involves accumulating XOR results with the negation of symmetrically positioned pairs of bits (0 or 1) and finding the maximum.
        We have observed that the minimum sampling rate for these methods is 5 MHz. As mentioned in Section \ref{sec_requirement}, different data rates have varying synchronization accuracy requirements. The lower the data rate, the larger the tolerable error, which allows for a lower sampling rate. Therefore, we can reduce the modulation order or increase redundant coding to lower the sampling rate, thus meeting the requirements of ultra-low-power applications.
        
\indent \noindent{\bf Resource Consumption:}
    Fig. \ref{eva_resource} shows the resource consumption of various methods with quantization on AGLN250 at different sampling rates. 
    We observe that SD\_Q and SA\_Q consistently consume the least resources at any sampling rate. 
    For instance, at a 5 MHz sampling rate, SD\_Q uses 853 D flip-flops, which is 8.5x fewer than NFT\_Q and 3.4x fewer than SST\_Q.
    Furthermore, the scalability of NFT\_Q and SST\_Q based on cross-correlation is poor, and their achievable additional functionalities are limited. 
    Specifically, at a 5 MHz sampling rate, NFT\_Q consumes 7208 D-flip-flops, which exceeds the AGLN250's total capacity of 6,144 D-flip-flops. This capacity must also accommodate modulation logic, leaving even less room for synchronization. Consequently, NFT\_Q cannot be deployed on the target hardware. Even if one attempts to operate it at a lower sampling rate to reduce resource consumption, Fig. \ref{eva_error_sr_Q} shows that at 1 MHz, its synchronization error (exceeding $8\mu s$) fails to meet the accuracy requirement. In contrast, SD\_Q and SA\_Q consume only 853 and 855 D-flip-flops, respectively, comfortably fitting within the budget. Thus, among the methods that meet the accuracy requirement, only the symmetry-based methods are feasible for practical low-power tags.

    In conclusion, taking into account sync accuracy and resource consumption under realistic hardware constraints, SD\_Q and SA\_Q based on mirror symmetry emerge as the superior synchronization strategies for low-power tags.

    \begin{figure*}[!t]
	\centering
	\begin{minipage}[b]{0.72\linewidth}
		\centering
		 \subfloat[Sync error.]{\includegraphics[width=0.31\linewidth]{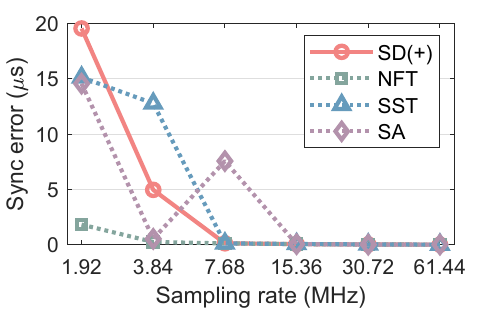}
			\label{eva_error_sr_simu}}
		\hfil
        \subfloat[Sync delay.]{\includegraphics[width=0.31\linewidth]{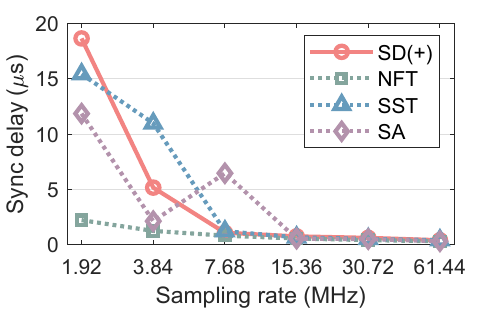}
                \label{eva_delay_sr_simu}}
		\hfil
	\subfloat[Total computational load.]{\includegraphics[width=0.31\linewidth]{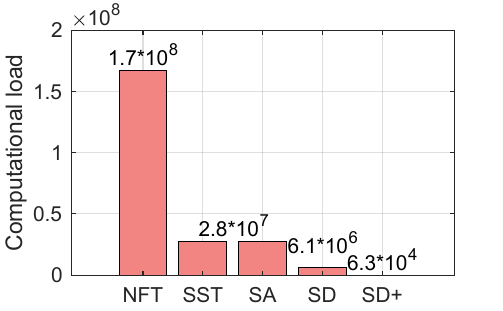}
			\label{eva_amount}}
	\caption{Sync error, delay, and computational load without quantization.}
        \label{eva_error_snr}
	\end{minipage}
  \hfil
     \begin{minipage}[b]{0.26\linewidth}
		\centering
		\centerline{\includegraphics[width=1\linewidth]{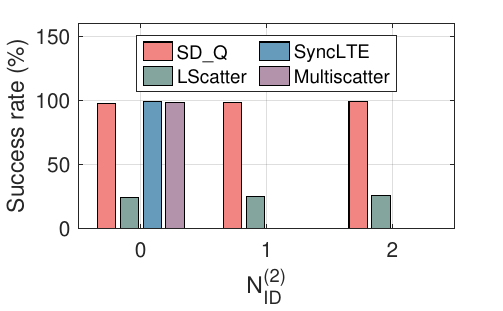}}
		\caption{Success rate.}
		\label{eva_success_rate}
	\end{minipage}
 \end{figure*}

\subsubsection{Comparison of Sync Methods without quantization}\label{sec_eva}
\ 
\newline
\indent \noindent{\bf Synchronization Accuracy and Delay:}
    As shown in Figures \ref{eva_error_sr_simu} and \ref{eva_delay_sr_simu}, we evaluate the synchronization accuracy and delay of the NFT, SST, SA, and SD(+) methods at various sampling rates. 
    Firstly, we observe that these methods require different minimum sampling rates to meet synchronization requirements. 
    The minimum required sampling rates for NFT, SST, SA, and SD(+) are 1.92 MHz, 7.68 MHz, 3.84 MHz, and 3.84 MHz, respectively. 
    The variation in minimum sampling rates is due to NFT having the longest template and hence requiring the lowest sampling rate. 
    In contrast, SST exhibits poor autocorrelation effects at lower sampling rates for symmetry determination, which worsens synchronization results with delayed symmetric positions, necessitating the highest sampling rate.
    Furthermore, we observe that for all synchronization methods, both synchronization error and delay decrease as the sampling rate increases. The minimum synchronization error and delay are 0.01 \textmu s and 0.3 \textmu s, respectively.  

    It is worth noting that at low sampling rates, SD exhibits higher synchronization errors than NFT and SST. This is expected because SD relies on envelope symmetry, which becomes more susceptible to noise when fewer samples are available per symbol. However, the critical comparison is not at an arbitrarily low sampling rate, but at the minimum sampling rate required to meet the $8\mu s$ accuracy bound. At these operating points, each method consumes vastly different hardware resources, as we analyze next.

     \begin{figure}[!t]
	\centering
	\begin{minipage}[b]{0.48\linewidth}
		\centering
		\centerline{\includegraphics[width=1\linewidth]{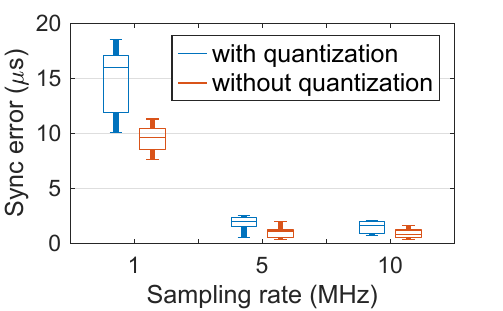}}
		\caption{Sync error with and without quantization.}
		\label{eva_error_quantization}
	\end{minipage}
  \hfil
     \begin{minipage}[b]{0.48\linewidth}
		\centering
		\centerline{\includegraphics[width=1\linewidth]{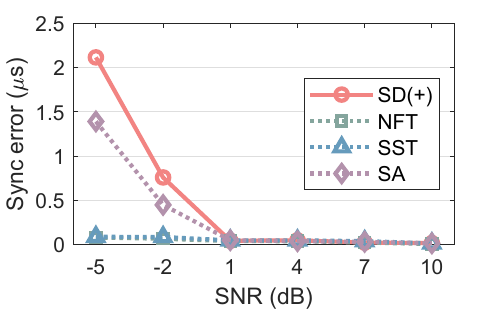}}
		\caption{Sync error across different SNRs.}
		\label{eva_error_snr}
	\end{minipage}
 \end{figure}
 
  \begin{table*}[]
 \centering
 \caption{Comparison of resource consumption at various sampling rates without quantization.}
	\label{tab_1}
  \scalebox{1.0}{
\begin{tabular}{cc|ccc|ccc|ccc}
\hline
\multicolumn{2}{c|}{\textbf{Sampling rate (MHz)}}                     & \multicolumn{3}{c|}{\textbf{1.92}}                                                       & \multicolumn{3}{c|}{\textbf{3.84}}                                                      & \multicolumn{3}{c}{\textbf{7.68}}                                                            \\ \hline
\multicolumn{2}{c|}{\textbf{Resource}}                                & Multiplier                 & Adder                      & D flip-flops                   & Multiplier                 & Adder                      & D flip-flops                  & Multiplier                   & Adder                        & D flip-flops                   \\ \hline
\multicolumn{1}{c|}{}                                  & \textbf{NFT} & {\color[HTML]{FE0000} 579} & {\color[HTML]{FE0000} 951} & {\color[HTML]{FE0000} 277,299} & 1,155                      & 1,911                      & 274,455                       & 2,307                        & 3,831                        & 1,147,767                      \\ \cline{2-11} 
\multicolumn{1}{c|}{}                                  & \textbf{SST} & 291                        & 471                        & 144,471                        & 579                        & 951                        & 287,799                       & {\color[HTML]{FE0000} 1,155} & {\color[HTML]{FE0000} 1,911} & {\color[HTML]{FE0000} 574,455} \\ \cline{2-11} 
\multicolumn{1}{c|}{}                                  & \textbf{SA}  & 97                         & 157                        & 48,157                         & {\color[HTML]{FE0000} 193} & {\color[HTML]{FE0000} 317} & {\color[HTML]{FE0000} 95,933} & 385                          & 637                          & 191,485                        \\ \cline{2-11} 
\multicolumn{1}{c|}{\multirow{-4}{*}{\textbf{Method}}} & \textbf{SD(+)}  & 0                          & 63                         & 1,575                          & {\color[HTML]{FE0000} 0}   & {\color[HTML]{FE0000} 127} & {\color[HTML]{FE0000} 3,175}  & 0                            & 255                          & 6,375                          \\ \hline
\end{tabular}}
\end{table*}

\begin{table*}[]
 \centering
 \caption{Power consumption for SD with and without quantization.}
	\label{tab_2}

 \scalebox{1.1}{
 \setlength{\tabcolsep}{4.2mm}{
\begin{tabular}{c|c|cccc|c}
\hline
\multirow{2}{*}{\textbf{Tag prototype}} & \multirow{2}{*}{\textbf{\begin{tabular}[c]{@{}c@{}}With/without \\  quantization\end{tabular}}} & \multicolumn{4}{c|}{\textbf{Component power consumption}}                                                       & \multirow{2}{*}{\textbf{Total}} \\ \cline{3-6}
                                        &                                                                                              & \multicolumn{1}{c|}{Digital core} & \multicolumn{1}{c|}{ADC}     & \multicolumn{1}{c|}{Oscillator} & Comparator &                                 \\ \hline
\multirow{2}{*}{\textbf{Verification}}  & With                                                                                         & \multicolumn{1}{c|}{0.51 mW}      & \multicolumn{1}{c|}{N/A}     & \multicolumn{1}{c|}{2.06 mW}    & 0.18 mW    & 2.75 mW                         \\ \cline{2-7} 
                                        & Without                                                                                      & \multicolumn{1}{c|}{0.57 mW}      & \multicolumn{1}{c|}{7.89 mW} & \multicolumn{1}{c|}{2.06 mW}    & N/A        & 10.52 mW                        \\ \hline
\multirow{2}{*}{\textbf{IC}}            & With                                                                                         & \multicolumn{1}{c|}{40.96 \textmu W}     & \multicolumn{1}{c|}{N/A}     & \multicolumn{1}{c|}{19.1 \textmu W}    & 10 \textmu W      & 70.06 \textmu W                        \\ \cline{2-7} 
                                        & Without                                                                                      & \multicolumn{1}{c|}{46.2 \textmu W}      & \multicolumn{1}{c|}{237 \textmu W}  & \multicolumn{1}{c|}{19.1 \textmu W}    & N/A        & 302.3 \textmu W                        \\ \hline
\end{tabular}}}
\end{table*}
    
\indent \noindent{\bf Resource Consumption:}
     We also evaluate the resource consumption of these synchronization methods at different sampling rates. 
    As shown in Table \ref{tab_1}, we observe that SD(+) exhibits the least resource consumption, falling below the resource capacity of AGLN250 at 6,144, whereas the other methods far exceed this capacity. 
    At the minimum sampling rate required to meet the accuracy target, NFT, SST, and SA consume 277,299, 574,455, and 95,933 D-flip-flops, respectively—all far exceeding the AGLN250's 6,144 D-flip-flop capacity. In contrast, SD(+) requires only 3,175 D-flip-flops (0 multipliers and 127 adders) at its required minimum sampling rate of 3.84 MHz, staying well within the budget. This represents a reduction of 87×, 181×, and 30× compared to NFT, SST, and SA, respectively. These results demonstrate that SD(+) is the only method that satisfies both the accuracy requirement and the hardware resource constraint, making it the sole viable candidate for practical deployment on low-power backscatter tags.

\indent \noindent{\bf Total Computational Load:}
    In addition, we evaluate the total computational load of the synchronization methods NFT, SST, SA, SD, and SD+.
    We use a sampling rate of 1.92 MHz and a total sample number N = 97,000, including 10 PSSs. 
    As shown in Fig. \ref{eva_amount}, the differential-based SD and SD+ consume significantly less total computational load than the other methods. For instance, the total computational load of NFT is 27x higher than that of SD. 
    Furthermore, specifically with the periodicity of PSS, the total computational load of SD+ is much lower than that of SD without PSS periodicity. The total computational load of SD+ is 63,000, which is 97x lower than that of SD.
    Note that the total computational load of SA and SST is similar since the time consumption of autocorrelation predominates in SST.

In summary, SD(+) using differential exhibits optimal comprehensive performance in terms of synchronization accuracy and resource consumption.
These results underscore that the advantage of our SD method lies not in outperforming others in raw accuracy across all sampling rates, but in its unique ability to deliver the required accuracy while respecting the severe resource limitations of low-power backscatter tags.

\subsection{Comparison with State-of-the-Art Synchronization Methods}
We further compare SD\_Q with synchronization methods adopted in existing backscatter systems. In our experiments, the sampling rate is set to 5 MHz. As shown in Fig. \ref{eva_success_rate}, SD\_Q consistently achieves an optimal synchronization success rate of 99\% across all cell IDs. In contrast, one-template-based methods such as SyncLTE and Multiscatter reach 99\% success rate only when \( N_{ID}^{(2)} = 0 \), but their performance drops to zero when \( N_{ID}^{(2)} = 1 \) or 2, as both rely on a template derived from the PSS with \( N_{ID}^{(2)} = 0 \). Additionally, LScatter, which detects the rising edge of the PSS, achieves a success rate of only around 25\% for all \( N_{ID}^{(2)} \) values, due to its inability to distinguish the PSS from other signals, resulting in significant synchronization errors. 
It is important to note that even when considering multi-template extensions of these methods—for example, NFT\_Q as the multi-template version of Multiscatter—they remain impractical for low-power tags. As shown in Fig. \ref{eva_resource}, NFT\_Q consumes 7208 D-flip-flops at 5 MHz, far exceeding the AGLN250's capacity of 6,144, whereas SD\_Q requires only 853 D-flip-flops. This further underscores that SD\_Q is the only method that simultaneously achieves high accuracy, universal applicability across cell IDs, and resource efficiency compatible with ultra-low-power hardware.

\subsection{Comparison of SD with and without quantization}
\subsubsection{Synchronization Accuracy}
    We evaluate the synchronization error of SD with and without quantization across sampling rates. 
    As shown in Fig. \ref{eva_error_quantization}, we observe that to achieve synchronization errors below 8 \textmu s, the minimum required sampling rate for both SD with and without quantization is 5 MHz, with median synchronization errors of 2 \textmu s and 1.5 \textmu s, respectively. 
    However, at 1 MHz sampling rate, the median synchronization errors for SD with and without quantization are 16 \textmu s and 9.5 \textmu s, respectively, exceeding 8 \textmu s.


\subsubsection{Power Consumption}
    We also evaluate the power consumption of SD with and without quantization in both verification and IC prototypes. 
    SD without quantization uses an ADC, while SD with quantization employs a comparator with a sampling rate of 5 MHz. 
    Table \ref{tab_2} illustrates that for both verification and IC prototypes, the total power consumption of SD with quantization is lower than that of SD without quantization. 
    Specifically, for the IC prototype, the power consumption of the digital core, oscillator, and comparator for SD with quantization is 40.96 \textmu W, 19.1 \textmu W, and 10 \textmu W respectively, resulting in a total power consumption of 70.06 \textmu W, which is 4.3x lower than that of SD without quantization.


\subsection{Impact of Different SNRs}
    Fig. \ref{eva_error_snr} shows the impact of different SNRs on synchronization errors. 
    We observe that in low SNR scenarios, cross-correlation-based methods like NFT and SST exhibit greater robustness than symmetry-based methods such as SA and SD(+). 
    At an SNR of -5 dB, both NFT and SST have synchronization errors of less than 0.1 \textmu s, which are 14x lower than SA and 21x lower than SD(+). 
    The primary reason is that SA and SD(+), designed for low power, are prone to symmetry disruption under low SNR conditions, whereas NFT and SST, based on cross-correlation, offer strong noise resistance.
    Our adaptive hybrid approach dynamically selects between SD\_Q and SST\_Q based on channel conditions, ensuring robust synchronization across diverse 5G backscatter scenarios.
   
\begin{figure} [t]
	\centering
	\begin{minipage}[b]{0.48\linewidth}
		\centering
		\centerline{\includegraphics[width=1\linewidth]{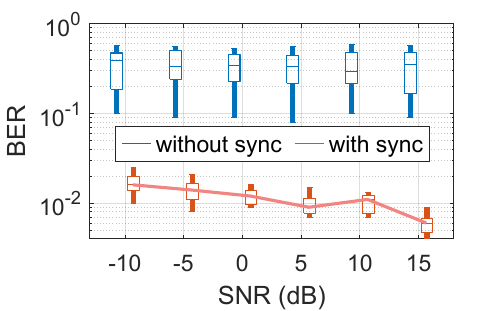}}
		\caption{BER across SNRs.}
		\label{eva_ber}
	\end{minipage}
 \hfil
	\begin{minipage}[b]{0.48\linewidth}
		\centering
		\centerline{\includegraphics[width=1\linewidth]{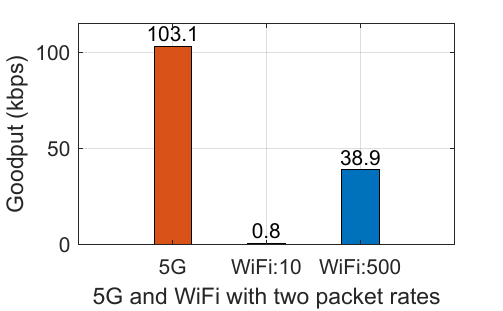}}
		\caption{Goodput.}
		\label{eva_wifi_nr}
	\end{minipage}
\end{figure}

\subsection{Application}
Finally, we evaluate the impact of synchronization on backscatter transmission across different SNRs. 
As shown in Fig. \ref{eva_ber}, we observe that at any SNR, the BER with SD synchronization is significantly lower than without synchronization, indicating the critical importance of SD synchronization for reliable end-to-end 5G backscatter communication. 
Specifically, at 15 dB SNR, the median BER with synchronization is 0.6\%, which is 55x lower than the BER without synchronization. 

Furthermore, we evaluate the goodput of 5G and WiFi. The WiFi packet rates are set at 10 and 500 pkts/s, with 10 pkts/s representing the typical rate for WiFi beacons. 
WiFi also employs single-symbol backscatter modulation. 
Fig. \ref{eva_wifi_nr} illustrates that 5G achieves a maximum goodput of 104 kbps, which is 135x higher than WiFi beacons.
This is because 5G traffic is continuous, whereas the WiFi signal is intermittent.

\section{Related Work}

\noindent{\bf Ambient Backscatter Communication:} 
    To reduce deployment costs, ambient backscatter technology has been proposed to enable communication using existing infrastructure \cite{gong2017towards,gong2016exploiting,chen2021reliable,xu2022enabling}.  
    Ambient backscatter initially used ambient TV signals as carriers \cite{liu2013ambient}. 
    Subsequently, many followers have employed various ambient signals for backscatter. 
    Mainstream indoor wireless technologies such as WiFi \cite{liu2021verification,liu2020vmscatter,yuan2023enabling,yuan2022subscatter}, Bluetooth \cite{jiang2023dances,jiang2023bidirectional}, and ZigBee \cite{li2018passive,xu2023bumblebee} support numerous connections for applications like smart homes. 
    LoRa \cite{jiang2021long,guo2022saiyan,peng2018plora} utilizes chirp spread spectrum (CSS) modulation to support long-distance backscatter transmission. 
    Cellular signals like LTE enable ubiquitous connectivity due to their continuity \cite{chi2020leveraging,feng2023heartbeating}. 
    5G, beyond continuity, supports high-speed and accessible backscatter due to its broadband width and frequency diversity.

\noindent{\bf Synchronization for Backscatter:} 
    Synchronization, as the core of communication, has been extensively studied in the literature. 
    Typical sync methods combine autocorrelation and cross-correlation to achieve the desired accuracy \cite{chen2020design,omri2019synchronization,nasraoui2014simply}. 
    For instance, WiFi employs autocorrelation using the periodicity of Short Training Field (STF) for packet detection and Long Training Field (LTF) cross-correlation for symbol alignment \cite{wang2007timing}. 
    Despite their high accuracy, these sync methods come with high computational complexity. 
    Hence, some researchers are focused on reducing complexity \cite{hu20185g,nassralla2015low,hsieh1999low}. 
    \cite{zhang2012low} proposes central-self-correlation (CSC)-based sync using the central symmetry of LTE PSS.
    However, these methods can lead to excessive resource consumption for backscatter. 
    To address these challenges, we apply differential based on PSS's unique mirror symmetry to reduce resource consumption.

\section{Discussion}
\noindent{\bf Practicality:} The gNodeB used in our experiments runs the open-source 5G stack OAI and emits compliant 5G signals over the air. The transmitted waveforms experience realistic propagation conditions including path loss and multipath, and are compatible with commercial 5G devices. In large-scale ambient deployments, the received SNR decreases with distance due to path loss, and our hybrid strategy is designed to adapt to such variations. In addition, bandwidth constraints do not affect the PSS symmetry, as the conjugate symmetry originates from the PSS sequence itself. 
More generally, our symmetry-based method can apply to cellular systems whose synchronization signals are constructed from structured sequences with favorable correlation properties, which are expected to remain in future cellular standards.

\section{Conclusion}
    We propose SD, a low-power and high-accuracy synchronization method for 5G backscatter. 
    Specifically, we apply differential techniques based on the unique mirror symmetry of the PSS envelope to convert the multiplication in cross-correlation into addition, significantly reducing computational resources without compromising accuracy. 
    Additionally, we introduce SD+, which further reduces total computational load by leveraging the periodicity of PSS.
    Extensive experiments demonstrate the superiority of SD in synchronization accuracy and resource efficiency. 
    We envision that SD, as an accurate and resource-efficient synchronization approach, can extend to other cellular networks such as 6G.

\section*{Acknowledgments}
This work was supported by NSFC Grant No. 62471451. The authors would like to thank Information Science Laboratory Center of USTC for the hardware/software services.



 

\bibliographystyle{IEEEtran}
\bibliography{ref}

\begin{IEEEbiography}[{\includegraphics[width=1in, clip,keepaspectratio]{figure/yunyunfeng.pdf}}]{Yunyun Feng (Student Member, IEEE)}
	received the Ph.D. degree from the School of Computer Science and Technology, University of Science and Technology of China in 2025, and is currently a postdoctoral researcher at the same institution. Her advisor is Prof. Wei Gong. Her research interests include backscatter communication and Internet-of-Things applications.
\end{IEEEbiography}

\begin{IEEEbiography}[{\includegraphics[width=1in,height=1.25in,clip,keepaspectratio]{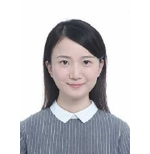}}]{Chenhong Cao}(Member, IEEE) received the B.S. and M.S. degrees in computer science from Northeastern University, China, in 2011 and 2013, respectively, and the PhD degree from Zhejiang University, in 2018. She is currently an associate researcher with the School of Computer Science and Technology, University of Science and Technology of China. Her research interests include the Internet of Things, network measurement, wireless and mobile computing.
\end{IEEEbiography}

\begin{IEEEbiography}[{\includegraphics[width=1in, clip,keepaspectratio]{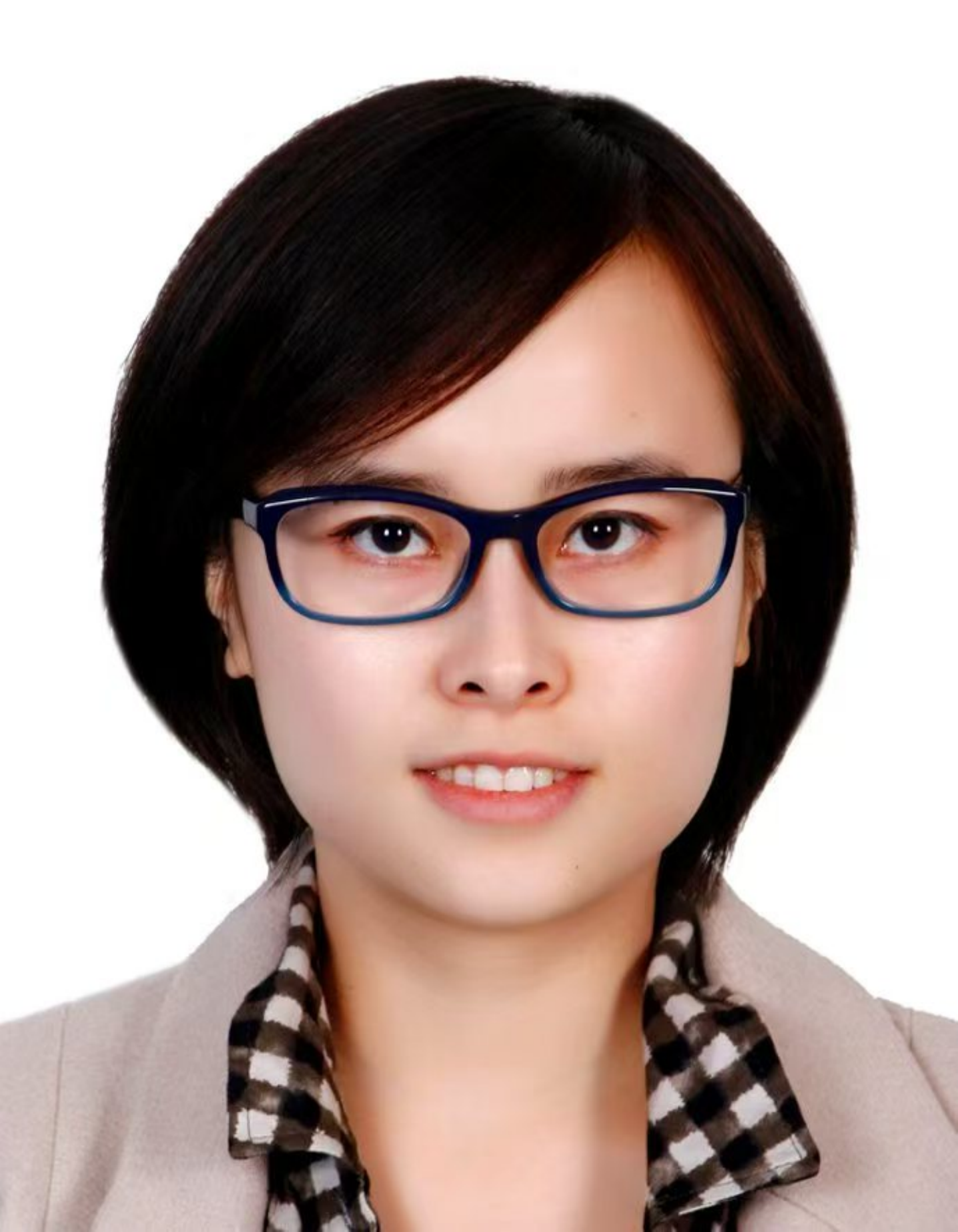}}]{Si Chen (Student Member, IEEE)}
	received the bachelor’s degree from the China University of Geosciences and the master’s degree from Simon Fraser University, where she is currently pursuing the Ph.D.
	degree with the School of Computing Science. Her
	research interests include wireless networks and big data.
\end{IEEEbiography}

\begin{IEEEbiography}[{\includegraphics[width=1in, clip,keepaspectratio]{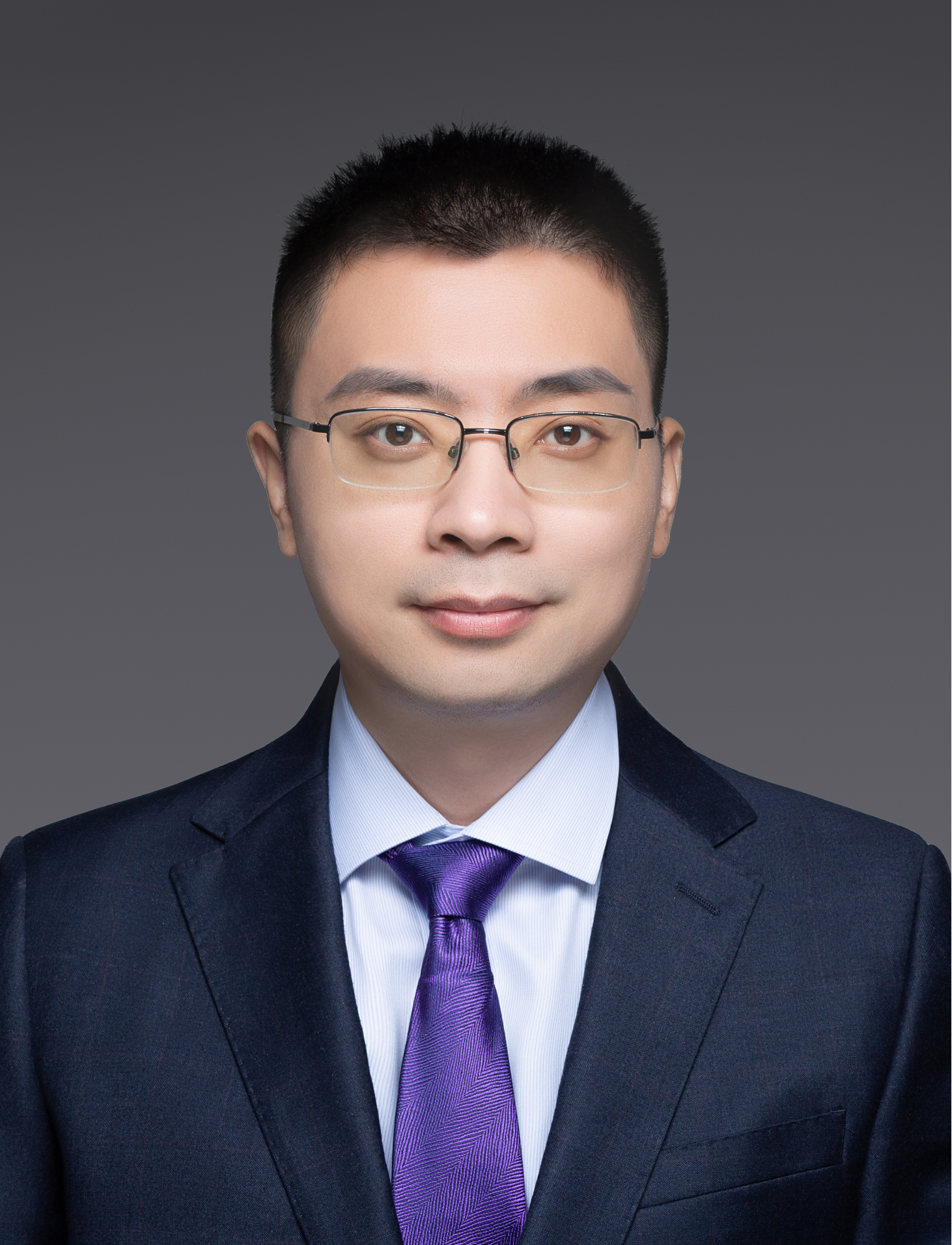}}]{Wei Gong (Senior Member, IEEE)}
	received the B.S. degree from the Department of Computer Science and Technology, Huazhong University of Science and
	Technology, and the M.S. and Ph.D. degrees from the School of Software and Department of Computer Science and Technology, Tsinghua University. 
	He is currently a Professor with the School of Computer Science and Technology, University of Science and Technology of China. 
	His research interests include backscatter networks, edge systems, and the IoT applications.
\end{IEEEbiography}

\vfill

\end{document}